\documentstyle[12pt]{article}
\def\Journal#1#2#3#4{#4 {#1} {\bf #2} #3}

\newcommand{\bm}[1]{\mbox{\boldmath $#1$}}

\def\CQG{\em Class. Quantum Grav.}
\def\JPA{\em J. Phys. A: Math. Gen.}
\def\PRD{\em Phys. Rev. D }

\def\IJT{\em Int. Jn. Theor. Phys.}

\def\Pary{\frac{\partial}{\partial y}}
\def\Parz{\frac{\partial}{\partial z}}
\def\Parx{\frac{\partial}{\partial x}}
\def\Part{\frac{\partial}{\partial t}}
\def\t{\dot{T_{1}^{}}}
\def\dt{\ddot{T_{1}^{}}}
\def\tt{\dot{T_{2}^{}}}
\def\dtt{\ddot{T_{2}^{}}}
\def\ttt{\dot{T_{3}^{}}}
\def\x{X_{1}'}
\def\dx{X_{1}''}
\def\xx{X_{2}'}
\def\dxx{X_{2}''}
\def\xxx{X_{3}'}
\def\dttt{\ddot{T_{3}^{}}}
\def\dxxx{X_{3}''}
\def\R{{\rm I\!R}}
\def\acc{\mbox{\cal{a}}}         
\def\d{\mbox{d}}

\def\wt{\widetilde}
\def\wh{\widehat}
\def\be{\begin{equation}}
\def\ee{\end{equation}}
\def\bea{\begin{eqnarray}}
\def\eea{\end{eqnarray}}
\def\bean{\begin{eqnarray*}}
\def\eean{\end{eqnarray*}}

\newtheorem{lemas}{Lemma}
\newenvironment{lema}{\begin{lemas}{\rm\hspace{-2mm}:\hspace{2mm}}}{\end{lemas}}

\newtheorem{teorem}{Theorem}

\newtheorem{conjetu}{Conjecture}

\newtheorem{corollary}{Corollary}[teorem]

\begin{document}

\title{$G_{2}$ COSMOLOGICAL MODELS SEPARABLE IN NON-COMOVING COORDINATES}

\author{Jos\'e M. M. Senovilla\thanks
{Also at Laboratori de F\'{\i}sica Matem\`atica, Institut d'Estudis Catalans,
Catalonia.}
\, and Ra\"ul Vera\footnotemark[1]\,\\
Departament de F\'{\i}sica Fonamental, Universitat de Barcelona,\\
Diagonal 647, E-08028 Barcelona, Catalonia, Spain.}

\maketitle
\begin{abstract}
We study new separable orthogonally transitive abelian $G_{2}$
on $S_{2}$ models with two mutually orthogonal integrable 
Killing vector fields. For this purpose we consider separability of the
metric functions in a coordinate system in which the velocity vector
field of the perfect fluid does not take its canonical form,
providing thereby solutions which are non-separable in comoving coordinates
in general. Some interesting general features concerning this class of
solutions are given. We provide a full classification for these models and
present several families of explicit solutions with their properties.
\end{abstract}

\section{Introduction}
The study of spatially inhomogeneous cosmological models has been of great
interest during past and present decades
\cite{KR,WW0,WW1,CCM,MCLLUM,ere94}.
The fact that the Universe is not 
exactly spatially homogeneous and the possibility of obtaining more general
solutions for very-early or late universe models are the main reasons for
such studies.
The spatially inhomogeneous models which have been studied systematically
are those
space-times admitting a maximally 2-dimensional group of local isometries
acting on spacelike surfaces (called $G_{2}$ on $S_{2}$).
Wainwright's classification \cite{WW1} for the abelian $G_{2}$ case
is a very useful tool in order to deal with such solutions. This
classification contains four sub-cases depending
on the features of the two Killing vector fields that generate the
group. The most simple sub-case, called B(ii), appears when the
Killing vectors are both hypersurface-orthogonal
(and hence the group acts orthogonally
transitively) and mutually orthogonal. This class with a perfect
fluid source has been already
treated under some extra conditions, as for example,
the imposition of additional homothetic or conformal
Killing vector fields (\cite{ali,MATHO} and references therein).
The most studied simplification
has been, however, the assumption of separability of the metric functions
in diagonal and canonical form \cite{WAGO,RUSE,AGGO}.

There have also been general studies on orthogonally transitive
$G_2$ cosmologies from a qualitative point of view, analyzing
the autonomous system of first-order partial differential equations
coming from the Einstein field equations by using methods from
the theory of dynamical sistems \cite{HEWWAIN,Llibre,RUSE}. The relations
between some of the known explicit solutions and these
theoretical studies are also analyzed in \cite{Llibre} and many references
therein.

It must be stressed that the majority of the explicitly known solutions
\cite{WAGO,RUSE,AGGO} have been obtained by means of the separability of
the metric functions in {\it comoving coordinates}, that is to say, such that
the velocity vector of the fluid takes its canonical form
$\bm{u} \propto \d x^0$. This involves {\it two} types of restrictions
because the separability of the metric functions is assumed in a
{\it particular} well-defined coordinate system. Thus, solutions which are
separable in other {\it non-comoving} coordinates have not been studied so far.
Notice that the choice of comoving coordinates will destroy in general any
{\it previously} assumed separability of the metric functions.
By the way, non-comoving coordinates have been already 
used before (a pioneering paper is \cite{McVitWilt}).

All this is fully explained in Sections 2 and 4. Section 2 is devoted to
presenting some general properties of the orthogonally transitive abelian
$G_{2}$ models, in particular the so-called ``coordinate interchange symmetry''
which allows to change the names of the non-ignorable coordinates for general
$G_2$ diagonal perfect-fluid spacetimes. In fact, this property holds also
for non-diagonal $G_2$ models (see \cite{MATHO}). Section 3 is devoted to
showing a broadly general property of spatially inhomogeneous perfect-fluid
models, that is, that they are in general extendible spacetimes. In other
words, solving the Einstein field equations for a perfect fluid one obtains
{\it regions} where the perfect-fluid character of the energy-momentum
tensor holds, but these regions may not (and in general they will not) be a
complete spacetime because the algebraic type of the energy-momentum tensor
cannot hold everywhere. This is a generic property of perfect-fluid
inhomogeneous spacetimes, and in particular it holds for the diagonal $G_2$
cases. This had not been remarked before precisely due to
the traditional use of comoving coordinates, because the comoving
coordinates hold obviously {\it only} at the true perfect-fluid region.
Actually, the use of non-comoving coordinates, as in the present
treatment, provides
natural extensions of the incomplete perfect-fluid regions whenever they are
extendible. These natural extensions keep the $G_2$ symmetry of the global
spacetime but, of course, the energy-momentum tensor cannot keep its
perfect-fluid form. The possible algebraic types allowed for these extensions
are also given in section 3.

The separation Ansatz in general coordinates is defined in section 4 in a
precise manner. This leads to a classification of the general separable models
depending on the values of two natural numbers related to the number of
linearly
independent functions appearing in the metric. The full classification is
presented and analyzed, including the explicit form of the field equations,
which now become simple systems of ordinary differential equations. The general
kinematical properties of the Weyl tensor, the Petrov type and other relevant
quantities are also given.

Finally, section 5 is devoted to the resolution and the study of the
properties of part of the cases classified in section 4. Several solutions
are explicitly given and many of them serve as illustrative examples of the
method and their general features. In particular, the coordinate interchange
symmetry is used to get physically adequate solutions. Furthermore, the
power of the method is manifested by presenting one of the solutions in its
canonical comoving coordinates, showing that it would have been very difficult
indeed to have found it in its comoving form despite the fact that it has
a perfectly simple expression in non-comoving coordinates. Some concluding
remarks are given at the end of the paper.

\section{$G_{2}$ cosmological models}
\label{sec:gen}

We are dealing with orthogonally transitive abelian $G_{2}$ models with
two mutually orthogonal integrable\footnote{By ``integrable'' vector
fields we mean hypersurface orthogonal vector fields.} Killing vectors.
The matter content is assumed to be a perfect fluid,
thus
there exists a time-like vector field $\vec u$
(velocity vector) such that 
the energy-momentum tensor takes the following form:
\be
T_{\alpha \beta}=(\rho +p)u_{\alpha}u_{\beta}+pg_{{\alpha}{\beta}},
\hspace{1cm}
u_{\alpha}u^{\alpha}=-1,
\label{decomp}
\ee
where $\rho$ and $p$ are the energy density and the pressure
of the fluid respectively.
The existence of two commuting integrable Killing vector fields
implies that the velocity
vector field is orthogonal to them and invariant under
the symmetry group. This, in turn, implies that $\vec u$ is integrable.
From this last result it can be proven a theorem (see
Wainwright \cite{WW1}) that assures the existence of {\em local\/}
coordinates $\{ x^{\alpha} \}$,
already adapted to both Killings, in which the metric takes a
diagonal form and such that $\vec u \propto \partial /
\partial x^0$ (canonical form of $\vec u$), so $\bm{u} \propto \d x^0$;
that is, there always exist coordinates {\em adapted\/} simultaneously
to the Killings
and the velocity vector so that the metric is diagonal.

Nevertheless, we can also choose {\em non-adapted coordinates\/}
to $\vec u$.
Of course, this will be relevant only when some restriction
has been imposed on the metric functions in a coordinate system.
Thus, the assumption
of separability allows us to distinguish two cases: the 
particular case when the coordinates that bring the metric functions in
separate form are chosen to be adapted to
the velocity vector, and the general one, when these coordinates are not
further restricted to be adapted to $\vec u$.
In this last generalized case, the velocity vector
field of the perfect fluid takes its most general form (restricted
to be integrable, orthogonal to the Killing vectors and
invariant under the $G_{2}$ group) in the coordinates that diagonalize the
metric and ``separate'' the metric functions.
The line--element for coordinates adapted to both
Killing vectors but not necessarily to the velocity vector field reads
\be
ds^{2}=-F_{0}^{2} \d t^{2} + F_{1}^{2} \d x^{2} + F_{2}^{} ( F_{3}^{2}
\d y^{2} + F_{3}^{-2} \d z^{2}) ,
\label{diag}
\ee
where  $F_{\alpha}=F_{\alpha}(t,x)$ and the Killings are
$\vec \xi=\partial /\partial y$, $\vec \eta =\partial / \partial z$,
while the velocity 1-form takes its most general form
\be
\bm{u}=u_0(t,x)\bm{\theta}^0+u_1(t,x)\bm{\theta}^1,
\label{u}
\ee
where $(u_0)^2-(u_1)^2=1$ in the orthonormal co-basis
\be
\bm{\theta}^0=F^{}_0 \d t,\hspace{1cm}
\bm{\theta}^1=F^{}_1 \d x,\hspace{1cm}
\bm{\theta}^2=F^{1/2}_2F^{}_3 \d y,\hspace{1cm}
\bm{\theta}^0=F^{1/2}_2F^{-1}_3 \d z.
\label{cobasis}
\ee

The non-zero components of the Einstein tensor
for the metric (\ref{diag}) in this frame are
$S_{00}$, $S_{01}$, $S_{11}$, $S_{22}$ and $S_{33}$, so the Einstein equations
for a perfect fluid become
\bea
S_{22}&=&S_{33}\label{ein1}\, ,\\
S_{01}^2&=&(S_{00}+S_{22})(S_{11}-S_{22})\label{ein2},
\eea
where it must be taken into account that the existence of the perfect fluid
is restricted to the region defined as follows:
\be
2S_{22}+S_{00}-S_{11}\not=0, \hspace{1cm}
\mbox{sign}(2S_{22}+S_{00}-S_{11})=\mbox{sign}(S_{00}+S_{22}).
\label{regio1}
\ee
Note that $\mbox{sign}(S_{00}+S_{22})=\mbox{sign}(S_{11}-S_{22})$ due to
(\ref{ein2}) whenever $S_{01}\neq 0$.
Equations (\ref{ein1}) and (\ref{ein2}) are the necessary conditions for
having two spacelike eigenvectors of the Einstein tensor with the same
eigenvalues ($p=S_{22}=S_{33}$) and a third eigenvector with the same
eigenvalue ($p$). Conditions (\ref{regio1}) assure both the existence
of this third eigenvector as well as its spacelike character. (We refer
to section \ref{sec:regions} for a deeper discussion on this topic.)

The perfect-fluid quantities defined in region (\ref{regio1}) are then 
\[
p=S_{22},\hspace{4em} \rho=S_{22}+S_{00}-S_{11},
\]
\be
(u_0)^{2}=\frac{S_{00}+S_{22}}{\rho + p},\hspace{3em}
(u_1)^{2}=\frac{S_{11}-S_{22}}{\rho + p},
\label{us1}
\ee
where the signs for $u_0$ and $u_1$ must be chosen such that the relation
\be
S_{01}=(\rho+p) u_{0}^{}u_{1}^{},
\label{s01}
\ee
which comes from (\ref{decomp}) and Einstein's field equations
in units with $8\pi G=c=1$, holds.
It still remains, of course, the freedom on the whole sign for $\vec u$,
which is usually chosen such that $u^{0}$ is positive (i.e. $u_0$ is negative).
This means that both $\vec u$ and $\partial / \partial x^0$ are
future--directed (say).

Let us introduce now an interesting property of 
coordinate interchange symmetry of the diagonal $G_2$
geometries which will be very useful for the purposes
of classification and the obtaining of solutions in the next
section.
In fact, this property can be generalized to non-diagonal $G_2$
B(i) \cite{WW1} models (see \cite{MATHO}).

\subsection*{Coordinate interchange symmetry}
Given any explicit function $f(t,x)$, let us define
\[
\tilde{f}(t,x)\equiv f(x,t).
\]
Upon this definition, we can construct a new line element 
(denoted by $\wh{ds} \mbox{}^{2}$) which will
consist of interchanging the $t$ and $x$ variables in the original
metric functions, that is 
\[
\wh{ds} \mbox{}^{2} =\wt{g_{\alpha\beta}} \d x^{\alpha} \d x^{\beta}.
\]
We can compute the Einstein tensor $\wh{S}_{\alpha\beta}$ of the new metric
$\wt{g_{\alpha\beta}}$.
It is then easy to find the relation between
$\wh{S}_{\alpha\beta}$ and $S_{\alpha\beta}$, which is simply:
\[ \begin{array}{lll}
\wh{S}_{22}=-\wt{S_{22}},  &  \wh{S}_{33}=-\wt{S_{33}}, &  \\
\wh{S}_{01}=\wt{S_{01}},   &  \wh{S}_{00}=\wt{S_{11}},  &
\wh{S}_{11}=\wt{S_{00}}.
\end{array}
\]
The first consequence of these relations is that the Einstein
equations (\ref{ein1}) and (\ref{ein2}) are invariant
under this coordinate interchange in the sense that the Einstein 
equations for the new line-element are
the old equations (\ref{ein1}) and (\ref{ein2}) with $t$ and
$x$ interchanged in the metric functions and, of course, in the derivative
operators (changing dots and primes).
In other words and explicitly
\bean
\wh{S}_{22}-\wh{S}_{33}=0  & \Longleftrightarrow  &
\wt{S_{22}}-\wt{S_{33}}=0, \\
(\wh{S}_{00}+\wh{S}_{22})(\wh{S}_{11}-\wh{S}_{22})=\wh{S}_{01}^{\;\;\;2}
&\Longleftrightarrow&
(\wt{S_{00}}+\wt{S_{22}})(\wt{S_{11}}-\wt{S_{22}})=\wt{S_{01}}\,^{2}.
\eean
In short, this means that if $ds^2$ is solution of equations
(\ref{ein1}) and (\ref{ein2}), so is $\wh{ds} \mbox{}^{2}$.
However, notice that the region of existence of the perfect fluid
for $\wh{ds} \mbox{}^{2}$, given by
\be
2\wh{S}_{22}+\wh{S}_{00}-\wh{S}_{11}\not=0, \hspace{1cm}
\mbox{sign}(2\wh{S}_{22}+\wh{S}_{00}-\wh{S}_{11})=
\mbox{sign}(\wh{S}_{00}+\wh{S}_{22}),
\label{regio2}
\ee
does not coincide in general neither with the previous region (\ref{regio1})
nor with the ``tilded'' region (\ref{regio1}) (with $t$ and $x$
interchanged). This region must be found separately in each particular
case. In fact, the {\em only\/} thing that can be said in general is that
the region where $2\wh{S}_{22}+\wh{S}_{00}-\wh{S}_{11}=0$ is the region
where $2S_{22}+S_{00}-S_{11}=0$ with $t$ and $x$ interchanged because
\[
2\wh{S}_{22}+\wh{S}_{00}-\wh{S}_{11}=
-(2\wt{S_{22}}+\wt{S_{00}}-\wt{S_{11}}).
\]
With regard to the fluid quantities, the coordinate interchange
transformation induces the following changes
\be
(\wh{u}_{0})^{2}=-\frac{\wt{S_{11}}-\wt{S_{22}}}{\wt{\rho} + \wt{p}}
\left(=``-\wt{({u^{1})^2}}\mbox{''}\right),
\hspace{1cm}
(\wh{u}^{1})^{2}=-\frac{\wt{S_{00}}+\wt{S_{22}}}{\wt{\rho} + \wt{p}}
\left(=``-\wt{({u^{0})^2}}\mbox{''}\right),
\label{us2}\ee
\[
\wh{\rho}=-\wt{\rho}, \hspace{1cm} \wh{p}=-\wt{p}.
\]
Note that in (\ref{us2}) there is no contradiction in the signs
of the quoted expressions
because they stand for their expressions (without the minus
sign) in (\ref{us1}), which are positive in the region (\ref{regio1}),
but change to be negative when they get transformed and defined in the
new region (\ref{regio2}).

\addtocounter{footnote}{1}
\section{A broadly general property of the $G_{2}$
solutions$\mbox{}^{\scriptscriptstyle{{\thefootnote}}}$}\label{sec:regions}
\footnotetext{We have entitled this section referring to the $G_2$
inhomogeneous models which are being treated in the present work,
but this property or similar ones apply to more general situations.}

We will assume that both Killing vector fields are globally defined
(well defined over the whole space-time manifold). If this did not happen,
we could take the open subset
of the original manifold admitting two spacelike isometries
as the manifold itself.
Notice that (\ref{regio1}) involves $t$ and $x$ in general, so that
(\ref{regio1}) restricts the allowed values of the coordinates
(apart from very particular cases which involve only the parametres
of the solution).
Then, formulae (\ref{regio1}) show that there appear different regions
for $G_2$ on $S_2$ spacetimes depending on the algebraic type
of the Einstein tensor at their points. Actually, this is a general
feature as will be shown in what follows.

Equations (\ref{ein1}) and (\ref{ein2}) restrict the Einstein
tensor, and hence the energy-momentum tensor via
the Einstein equations, to have 
four possible algebraic types:
\{1,(111)\}, \{(1,11)1\}, \{(1,111)\}, and \{(2,11)\} in Segr\'e's
notation (see \cite{KRAM} and references therein). Only the first three types
admit a timelike eigenvector (which is unique only in the first case),
while in the fourth type there exists a null eigendirection.
The explicit canonical form of $S_{\alpha\beta}$ for each Segr\'e
type together with the regions where the respective types hold
are given below:  
\begin{description}
\item[a)] $S_{\alpha \beta}=\rho \, v_{\alpha}^{0}v_{\beta}^{0} + 
p\,( v_{\alpha}^{1}
v_{\beta}^{1}+ v_{\alpha}^{2}v_{\beta}^{2} +v_{\alpha}^{3}v_{\beta}^{3})$,
at the region defined by
\bean
\{2S_{22}+S_{00}-S_{11}\not=0,\;
\mbox{sign}(2S_{22}+S_{00}-S_{11})=\mbox{sign}(S_{00}+S_{22})\},
\eean
\item[b)] $S_{\alpha \beta}=-a\,w_{\alpha}^{0}w_{\beta}^{0}+b\,
w_{\alpha}^{1}w_{\beta}^{1}+a\,(w_{\alpha}^{2}w_{\beta}^{2}+w_{\alpha}^{3}
w_{\beta}^{3})$, at the region given by
\bean
\{2S_{22}+S_{00}-S_{11}\not=0,\;
\mbox{sign}(2S_{22}+S_{00}-S_{11})=-\mbox{sign}(S_{00}+S_{22})\},
\eean
\item[ci)] $S_{\alpha \beta} \propto g_{\alpha \beta},\hspace{3mm}
\mbox{at the region with }\{2S_{22}+S_{00}-S_{11}=0,\;
S_{01}=0\}$,
\item[cii)] $S_{\alpha \beta} = c \, k_{\alpha} k_{\beta}-d \, 
g_{\alpha \beta},\hspace{.5cm}
(k_{\alpha}k^{\alpha}=0,\mbox{ } c\not=0)$, at the region
\bean
\{2S_{22}+S_{00}-S_{11}=0,\;
S_{01}\not=0\},
\eean
\end{description}
where both $\{\mbox{\boldmath $v$}^{\alpha}\}$ and
$\{\mbox{\boldmath $w$}^{\alpha}\}$
are orthonormal cobases.
The case a) corresponds to the perfect fluid region which we will call
the $\cal A$-region from now on. The zone where the case b) holds will
be called the $\cal B$-region.
Here, there is a {\it timelike} eigenvector of $S_{\alpha\beta}$
with eigenvalue equal to $p\equiv S_{22}=S_{33}$ (in $\cal A$, all three
eigenvectors with eigenvalue $p$ are spacelike).
The region defined by $\, 2S_{22}+S_{00}-S_{11}=0 \,$ is the border $\cal F$
(cases ci) and cii)), that divides the space-time manifold in the
$\cal A$ and $\cal B$ regions in the sense that for every continuous curve
containing points in $\cal A$ and $\cal B$, there always exists at least
one point of the curve in $\cal F$ (this is why we call it a border;
see also \cite{PRHMc}).
The assumption of analyticity of the Einstein tensor
on the whole
manifold assures that $\cal F$ is either the entire spacetime or
its interior is empty in the manifold topology,
while the mere assumption of smoothness implies that 
the $\cal A$ and $\cal B$-regions are open sets.

The behaviour of the fluid velocity vector (defined in $\cal A$)
when approaching the border $\cal F$ varies
depending on whether case ci) or cii) holds at $\cal F$ (this can be
easily seen from (\ref{us1})
and (\ref{ein2}), or also from (\ref{s01})):
if $S_{01}\not= 0$ at $\cal F$, the components of $\vec u$ in the orthonormal
co-basis must diverge when approaching $\cal F$, but this does not necessarily
happen when $S_{01}=0$ at points of $\cal F$.
Then, in a general situation there will be parts of the
border $\cal F$ where $\vec u$ diverges (see \cite{hallrendall} for a general
discussion).

Therefore, in general, a space-time corresponding to a solution 
of equations
(\ref{ein1}) and (\ref{ein2}) will be divided into three regions 
(not necessarily
connected) depending on the Segr\'e type of the energy-momentum tensor
at their points. This could be somewhat expected due to the quite general
conditions we are using. 
Our main interest, of course, will be the perfect-fluid region
$\cal A$, where we can construct a coordinate system in which
$\vec u$ takes its canonical (comoving) form while keeping the diagonal form
of the metric (see the previous section \ref{sec:gen}).
These coordinates are only defined
in this region in general, because the change of coordinates is not 
valid where $\vec u$ diverges, that is, on $\cal F$,
so the search of perfect-fluid solutions in comoving coordinates
may lead to solutions with coordinate singularities
that would correspond to an $\cal F$ border. 

Therefore, it seems a natural feature of the $G_{2}$ perfect-fluid
solutions to be extendible across $\cal F$, and the extensions
cannot keep the perfect-fluid character over the whole spacetime.
Of course, the above problem has a solution if we do not use comoving
coordinates, which is our purpose in this paper. By using non-comoving
coordinates and solving Einstein's equations (\ref{ein1}) and (\ref{ein2})
we obtain, in one single stroke, both regions $\cal A$ and $\cal B$ and the
border $\cal F$. Thus, by using this method of obtaining perfect-fluid
solutions we also get extensions of the $\cal A$ regions, that is, of the
perfect-fluid regions which are extendible across the border $\cal F$
where the (comoving) coordinate singularity appears.

\section{Separability in general: A classification for the general
models}
\label{sepgen}
Due to the diagonal form of the metric (\ref{diag}), the notion of
separability will be applied to the metric functions in the obvious
way:
\[
F_{\alpha}(t,x)={\cal T_{\alpha}}(t){\cal X_{\alpha}}(x).
\]
Once this is assumed, two different cases appear: 
(i) $t$,$x$ adapted to $\vec u$ (comoving coordinates)
and (ii) $t$,$x$ non-adapted to $\vec u$ (non-comoving coordinates).

All the possible solutions in case (i) have been already identified
and studied in \cite{WAGO,RUSE,AGGO}. Ref. \cite{WAGO} finds the
general solution
under the extra assumption $F_{3}=F_{3}(t)$ (see (\ref{diag}))
which implies that the three-slices orthogonal to the fluid congruence
are conformaly flat. In ref.\cite{RUSE}, the remaining solutions are
identified unless in the special case of $p=\rho$. Finally,
\cite{AGGO}, provides the $p=\rho$ solutions which complete the
comoving case. In this paper, we study the second (non-comoving) case.

With the help of separability we get the following line element
\be
ds^{2}=e^{2f_{1}} (-\d t^{2} + \d x^{2}) + e^{f_{2}}
( e^{2f_{3}} \d y^{2} + e^{-2f_{3}} \d z^{2}),
\label{ds2}
\ee
where $f_{a}(t,x)=T_{a}(t)+X_{a}(x)$ $(a=1,2,3)$,
while $\bm{u}$ takes its most general (non-canonical) form (\ref{u}).
The explicit components of the Einstein tensor in the orthogonal frame
(\ref{cobasis}) are given in Appendix \ref{ap:deffun}.
Equations (\ref{ein1}) and (\ref{ein2}) read then
\be
\dttt +\tt \ttt=\dxxx + \xx \xxx=\mbox{K},
\label{AA}
\ee
\be
\left( \t \xx + \tt \left( \x - \frac{1}{2} \xx\right) -2 \ttt \xxx
\right) ^{2}=
(M_{7}^{}(t)+ N_{8}^{}(x))(M_{8}^{}(t)+ N_{7}^{}(x)),
\label{expli}
\ee
where the dots and primes denote derivatives with respect to $t$ and $x$
respectively, $\mbox{K}$ is a separation constant, and the $M(t)$'s
and $N(x)$'s
functions stand for definite combinations of the first and second
derivatives of the metric functions which are explicitly defined in Appendix
\ref{ap:deffun}. Equation (\ref{expli}) can be re-written in the form
(see Appendix \ref{ap:deffun})
\be
\sum_{i=1}^{8}M_{i}(t)N_{i}(x)+M_{7}(t)M_{8}(t)+N_{7}(x)N_{8}(x)=0,
\label{expli2}
\ee
and differentiating this last expression with respect to $t$ and $x$
we get the following equation:
\be
\sum_{i=1}^{8} \dot{M}_{i}^{} N'_{i}=0.
\label{bo}
\ee 
This implies that if we define $n$ and $q$ to be the number of
linearly independent
functions among the sets $\{\dot{M}_i^{}\}$ and $\{N'_i\}$ respectively,
we have $q\leq 8-n$, that is,
there are, at most, $8-n$ linearly independent functions among
the \{$N'_{i}$\}.
Thus, at first sight, $n$ might take all values from 0 to 8, but
this {\em will not be necessary\/} eventually because of the
property of coordinate interchange symmetry
we have shown previously in section \ref{sec:gen}:
thanks to the $t\leftrightarrow x$ ``symmetry'', we have a way to
relate solutions of equations (\ref{AA}) and (\ref{expli})
interchanging their $n$ and $q$ numbers. Thus, if a solution has $n=3$
and $q=5$,
for instance, then by using the coordinate interchange we get another solution
with $q=3$ and $n=5$, and so on.
Therefore we must only treat the Einstein equations for the values
of $n$ ranging from 0 to 4. The rest of cases can be studied
by means of the coordinate interchange.

Notice however that, in fact, we have only three original functions of $t$,
\{$T_{a}$\}. Therefore, we can define another integer $m$ as the number
of linearly independent functions among the \{$T_{a}$\} (obviously $m$ runs 
from 1 to 3 because
$m=0$ avoids any dependence on $t$ of the metric). Of course, $m$
will be related with the previously defined $n$.
The combination of both $n$ and $m$ gives us a way of obtaining solutions
and a classification for them.

To proceed with this classification we will start with any value
of $m$, and then this $m$ is related on each case (using some results
given in the Appendix \ref{ap:lemes})
with the number of linearly independent functions among the set
\{$1,M_{i}$\}, which is exactly $n+1$ due to Lemmas 1 and 2 of
Appendix \ref{ap:lemes}.
Then, each of the three cases $m=1,2,3$ will be divided into
the possible values that $n$ can take for each case.
For a given pair
$\{m,n\}$ some subdivisions may appear depending on the
posible relations between the set of functions \{$1,M_{i}$\}.
These different possibilities will be given below.
Nevertheless, at this stage, we
have not used yet the equations (\ref{AA}) and (\ref{expli}) explicitly,
from where new relations arise. These new relations together with the
previous ones provide the systems of differential equations for the
$t$-functions and for the $x$-functions, from where new
restrictions may appear
after their compatibilization (if necessary).
Therefore, the aim of the next subsections is to give the steps
in dividing the cases for a given $m$ (subsections \ref{m1},
\ref{m2} and \ref{m3}). Then, in subsetion \ref{com}, equations
(\ref{AA}) and (\ref{expli}) are treated in order to
present the complete set of equations for the $t$- and $x-$functions. 
All throughout the next subsections, the latin lower-case characters
will stand for constants.

\subsection{Case $m=1$}
\label{m1}
We have for this case $T_{a}(t)=c_{a} T(t)$
and we need to impose
$T(t)\not\equiv 0$ and at least a non-vanishing
$c_{a}$ to assure one linearly independent function among the \{$T_{a}$\}.
Furthermore, $\dot{T}$ will be non-zero because otherwise
the metric would be static.
It is very easy now to establish the possible values of $n$
because, for $L$ running from 1 to 6, we have $M_{L}\propto
\dot{T}^{2}$, and the only
other possible independent function contained in \{$M_{i}$\}
is $\ddot{T}$, appearing in $M_{7}$ and $M_{8}$.
Therefore there are at most 3 linearly independent functions
involved in \{$1,M_{i}$\}
and, consequently, $n$ can take the values
0,1, and 2,
dividing the case $m=1$ into three subclasses.
We give now the characterization of these subclasses for
a given $n$ that will be used for the full analisys of the case $m=1$
in Section \ref{fullm1}.

\begin{description}
\item[I. $n=0$ :] this means ${\dot M}_{i}=0$ for ($i= 1\ldots 8$),
which is equivalent to $\dot T=1$ (the constants are absorved by the
$c_{a}$ coefficients).

\item[II. $n=1$ :] to avoid the previous case
the two linearly independent functions must be chosen as
\{$1,\dot{T}^2$\} and we must also have
$M_{7}=c_{71}\dot{T}^2+b_{7}$ and $M_{8}=c_{81}\dot{T}^2+b_{8}$.
Nevertheless, $b_{7}$ and
$b_{8}$ can be absorved by $M_{7}$ and $M_{8}$ (see Appendix \ref{ap:deffun}).
This case, taking into account the $t$-equation in
(\ref{AA})
and the fact that there is at least a non-vanishing $c_{a}$, implies
an equation for $T(t)$ with the following form: $\ddot{T}=a\dot{T}^2+b$.

\item[III. $n=2$ :] now, there is a relation of the form
$a+b\dot{T}^2+cM_{7}+dM_{8}=0$ and
there appear two different possibilities depending on wether
$c\not= 0$ or $c=0$.
\begin{description}
\item[(i)] $c\not= 0$. Then we can take \{$1,\dot{T}^2,M_{8}$\} as
linearly independent functions, and
$M_{7}=c_{71}\dot{T}^2+c_{72} M_{8}$.
\item[(ii)] $c=0 (\Rightarrow d\not= 0)$ Then
\{$1,\dot{T}^2,M_{7}$\} can be chosen as linearly independent functions, and
$M_{8}=c_{81}\dot{T}^2$.
\end{description}
Again, the possible additional constants
have been absorved in $M_{7}$ and $M_{8}$ as in the previous case.
\end{description}

\subsection{Case $m=2$}
\label{m2}
In this case we have $T_{a}(t)=c_{a}T(t)+d_{a}K(t)$
where \{$T(t),K(t)$\} are two linearly independent functions. We also
need to impose, of course, that the matrix
composed by $c_{a}$'s and $d_{a}$'s has rank two. This case can also
be treated dealing directly with the \{$T_{a}$\} functions and dividing
this class depending of which pair is taken to be linearly independent,
but we have preferred to introduce the functions $T(t)$ and
$K(t)$ for the sake of compactness and brevity.  

In fact, $1,T(t),K(t)$ can be assumed to be three
linearly independent functions,
as otherwise we could reduce this class to the case $m=1$. For, suppose, on
the contrary, that there existed a linear relation
$aT(t)+bK(t)+c=0$. Then, if $a=0$ we would have that $K(t)$ is a constant
and therefore, it could be set equal to zero because the terms $e^{d_{a}K}$
in the metric can be absorved into the coordinates. If $a\not=0$
we would have $T(t)=-(b/a)K(t)-c/a$, thus
$T_{a}(t)=(-c_{a}b/a+d_{a})K(t)-c_{a}c/a$,
and again, redefining the constants and absorving the constant terms into
the coordinates, we could set $T(t)\equiv 0$, in contradiction.
From Lemma 1 it follows then that $\dot T$ and $\dot K$
are two linearly independent functions, and therefore, Lemma 3 of the
Appendix \ref{ap:lemes} implies
that the set
\{$\dot{T}^2,\dot{K}^2,\dot{T}\dot{K}$\} consists of 3 linearly
independent functions.
This, in turn, means that among \{$M_{L}$\} ($L=1\ldots 6$) there are
exactly 3 linearly independent functions
(see Appendix \ref{ap:lemes}).

In $M_{7}$ and $M_{8}$ there appear two other functions ($\ddot{T}$ and
$\ddot{K}$) that can be linearly independent from the rest. Therefore,
in the set \{$1,M_{i}$\} we have a minimum of 3 linearly independent
functions and a maximum of 6 so that $n$
can take the values 2,3,4, and 5, but this last case ($n=5$) does not need
to be treated thanks to the coordinate interchange symmetry
$t\leftrightarrow x$.
In summary, the case $m=2$
is divided into three subclasses $n=2$, $n=3$ and $n=4$.

\subsection{Case $m=3$}
\label{m3}
In this case the three functions \{$T_{a}$\} are linearly independent.
Nevertheless, we will use here three generic independent functions
in order to give a more compact presentation of the cases. Thus,
let $T(t),K(t),Q(t)$ be
three linearly independent functions such that
$T_{a}(t)=c_{a}T(t)+d_{a}K(t)+e_{a}Q(t)$,
where the determinant of the 3x3 matrix composed by the constants must
be non-zero.

Analogously to what is explained in the second paragraph of
subsection \ref{m2}, the set \{$1,T(t),K(t),Q(t)$\} consists of four 
linearly independent functions, as otherwise we could reduce this
class to the case $m=2$.
From Lemma 1 it follows then that $\dot{T}$, $\dot{K}$, and $\dot{Q}$
are three linearly independent functions, and Theorem 1
implies then that there are at least 5 linearly independent functions
among the set
\{$\dot{T}^2,\dot{K}^2,\dot{Q}^2,\dot{T}\dot{K},
\dot{T}\dot{Q},\dot{K}\dot{Q}$\},
that is, among the $M_{L}$
functions, and hence, among \{$1,M_{i}$\}. Therefore, $n$ can take the
values from 4 to 8, so that only the case $n=4$ has to be treated.

	The three cases with $m=2$ and the single case with $m=3$ will
not be solved explicitly in this paper. Nevertheless, we present the
whole set of equations and the kinematical quantities {\it in general} in
the next two subsections.

\subsection{The complete set of equations}
\label{com}
Now, we proceed with the obtention of the complete set
of equations coming from equations (\ref{AA}) and (\ref{expli})
once we have chosen the pair $\{m,n\}$. As was
explained above, these equations together with the systems given in
the previous
subsections will form the full set of equations for the $t$- and $x$-
functions.
In fact, equations (\ref{AA}) need no further treatment and will not
be repeated in this subsection. Thus, we focus on
equation (\ref{expli}) written in its form (\ref{expli2}).

We choose $\{1,m_A(t)\}\, (A=1 \ldots n)$ to be the $n+1$ given linearly
independent functions such that
\be
M_i(t)=\sum^n_{A=1}c_{iA}m_A(t)+b_i \hspace{1cm} (i=1 \ldots 8),
\label{Mi}
\ee
where the constants $c_{iA}$ form a matrix of rang $n$.\footnote{Note
that we could set $b_7=b_8=0$ without loss of generality
(see Appendix \ref{ap:deffun}), but this will not be always used.}
Therefore, we take the functions $m_A(t)$ and the constants $b_i$
to be the fundamental objects with regard to the $M_i$ functions.
Now, from (\ref{bo}), using (\ref{Mi}) and the fact that
$\{\dot{m}_A(t)\}$ are lineraly independent (Lemma 1), there appear $n$
linearly independent relations between $N'_i$. This implies
that there are at most $8-n$ linearly independent
functions in $\{N'_i\}$. Thus, we choose $\{1,n_B(x)\}\, (B=1 \ldots (8-n))$
to be the $9-n\equiv (8-n)+1$ linearly independent functions such that
\be
N_i(x)=\sum^{8-n}_{B=1}d_{iB}n_B(x)+a_i  \hspace{1cm} (i=1 \ldots 8),
\label{Ni}
\ee
where the constants $d_{iB}$ have no restriction a priori.

Now, derivating equation (\ref{expli2}) with respect to $x$ 
and using (\ref{Mi}) we get
\[
\sum^n_{A=1}\left(\sum^8_{i=1}c_{iA}N'_i\right)m_A(t)
+\sum^8_{i=1}b_iN'_i+
\left(N'_7 N'_8\right)'=0.
\]
Using the fact that $\{1,m_A(t)\}$ are linearly independent functions,
from this last expression it follows that
$\sum^8_ic_{iA}N'_i=0$ and 
$\sum^8_ib_iN'_i+\left(N_7 N_8\right)'=0$.
Integrating this pair of expressions we get
\bea
&&\sum^8_{i=1}c_{iA}N_i={\cal C}_A \hspace{1cm} (A=1 \ldots n),
\label{Ca}\\
&&\sum^8_{i=1}b_iN_i+N_7 N_8=B,
\label{B}
\eea
where $\cal{C}_A$ and $B$ are constants.
Differentiating equation (\ref{expli}) with respect to $t$
and using (\ref{Ni}) we obtain the corresponding expressions for the $M_i$
functions. The equation analogous to (\ref{Ca}) does not
give relevant information (only relations between constants
that will not be used at the end), while the analogous to (\ref{B})
reads
\be
\sum^8_{i=1} a_i M_i + M_7 M_8=A.
\label{A}
\ee
We take now  the original Eq.(\ref{expli2}) and substitute the functions $M_i$
from (\ref{Mi}) and the terms $M_7 M_8$
and $N_7 N_8$ isolated from (\ref{A}) and (\ref{B}) respectively.
Using again the fact that $\{1,m_{A}(t)\}$ are lineraly independent
and taking (\ref{Ca}) into account, the following relations arise
\[
\sum^8_{i=1}a_i\; b_i=A+B,\hspace{1cm}
\sum^8_{i=1}a_i\; c_{iA}={\cal C}_A.
\]
Putting these last expressions into (\ref{A}) we finally get
\be
\sum_{A=1}^{n}{\cal{C}}_A m_A+M_7 M_8+B=0.
\label{m7m8}
\ee
Thus, equation (\ref{expli2}) splits into the equivalent
set of {\it ordinary} differential equations (\ref{Ca}),
(\ref{B}), and (\ref{m7m8}).
Therefore, the complete set of equations consists of
(\ref{AA}),(\ref{Ca}),(\ref{B}), and (\ref{m7m8}) together with
the relations between the $t$-functions arising from the classification
of the previous subsections.

\subsection{Kinematical quantities and the Weyl tensor}
\label{Kine}
For the sake of compactness in the expressions of the kinematical properties
of $\vec u$ we define first the following objects
\bean
w_{0}^{}&\equiv&e^{2f_{1}} \left( S_{00}+S_{22} \right)=M_{7}(t)+N_{8}(x),\\
w_{1}^{}&\equiv&e^{2f_{1}} \left( S_{11}-S_{22} \right)=M_{8}(t)+N_{7}(x),\\
\Sigma&\equiv&e^{2f_{1}} S_{01},
\eean
so the Einstein equation (\ref{ein2}) reads simply $w_{0}^{}w_{1}^{}=
\Sigma^{2}$,
and the perfect-fluid quantities take then the following form
\bean
\rho+p=e^{-2f_{1}}(w_{0}^{}-w_{1}^{}),&\hspace{1cm}&
\rho-p=e^{-2f_{1}}(\ddot{T}_{2}^{}+\tt^{2}-X''_{2}-\xx^2),\\
(u_{0}^{})^{2}=\frac{w_{0}^{}}{w_{0}^{}-w_{1}^{}},&\hspace{1cm}&
(u_{1}^{})^{2}=\frac{w_{1}^{}}{w_{0}^{}-w_{1}^{}}.
\eean
The $\cal A$-region defined in section \ref{sec:regions} is given now
by the conditions $w_0-w_1\neq 0$, $\mbox{sign}(w^{}_0-w^{}_1)
=\mbox{sign}(w^{}_0)$, which can be combined with the extra
requirement $\rho+p>0$ to give
\be
w^{}_0>0, \hspace{2cm}w^{}_0-w^{}_1>0,
\label{cond}
\ee
which become the necessary and sufficient conditions to have a
perfect-fluid source satisfying the energy condition $\rho+p>0$.
From now on, the domain defined by these conditions (\ref{cond})
will be referred to as the ${\cal A}_E$-region.
The relative signs of $u^{}_0$ and $u^{}_1$ are determined then by $\Sigma=
(w^{}_0-w^{}_1)u^{}_0 u^{}_1$.

The fact that $\vec u$ is integrable implies that its vorticity
vanishes, that is,
$\omega_{\alpha\beta}=0$. The expansion reads as follows,
\bean
\theta =\frac{e^{-f_{1}}}{u_{0}^{}(w_{0}^{}-w_{1}^{})^{2}} \left\{
(w_{0}^{}-w_{1}^{}) \left[\Sigma \left(\x+\xx\right)-w_{0}^{}
\left( \t+\tt\right)\right]\right.
\hspace{15mm} \\
\left.+w_{0}^{}\Sigma'-w'_{0}\Sigma+\frac{1}{2}(\dot{w}_{0}^{}w_{1}^{}-
w_{0}^{}\dot{w}_{1}^{}) \right\}.
\eean
The non-zero components of the acceleration computed in the co-basis
given in (\ref{cobasis}) are
\bean
&&\acc_{0}^{}=\frac{u_{1}^{}}{u_{0}^{}}\acc_{1}^{},\\
&&\acc_{1}^{}=-\frac{e^{-f_{1}}}{(w_{0}^{}-w_{1}^{})^{2}}
\left\{ (w_{0}^{}-w_{1}^{}) \left( \Sigma \t -w_{0}^{} \x \right) \right.
\hspace{6cm}\\
&&\hspace{6cm}
\left.+w_{0}^{}\dot{\Sigma}-\dot{w}_{0}^{}\Sigma+\frac{1}{2}(w'_{0}w_{1}^{}-
w_{0}^{} w'_{1}) \right\},
\eean   
while the shear tensor has the following non-zero
components
\bean
&&\sigma_{00}^{}=\left( \frac{u_{1}}{u_{0}} \right)^{2} \sigma_{11}^{},\\
&&\sigma_{01}^{}= \frac{u_{1}}{u_{0}} \sigma_{11}^{},\\
&&\sigma_{11}^{}=-e^{-f_{1}} \frac{u_{0}^{}}{(w_{0}^{}-w_{1}^{})}
\left( \Sigma \xx-w_{0}^{}\tt \right)+\frac{2}{3}(u_{0}^{})^{2}\theta,\\
&&\sigma_{22}^{}=-\frac{e^{-f_{1}}}{u_{0}^{}(w_{0}^{}-w_{1}^{})}
\left( w_{0}^{}\ttt-\Sigma\xxx \right)+\frac{1}{2}\frac{1}{(u_{0}^{})^{2}}
\sigma_{11}^{},\\
&&\sigma_{33}^{}=\frac{e^{-f_{1}}}{u_{0}^{}(w_{0}^{}-w_{1}^{})}
\left( w_{0}^{}\ttt-\Sigma\xxx \right)+\frac{1}{2}\frac{1}{(u_{0}^{})^{2}}
\sigma_{11}^{},
\eean
and the shear scalar $2\sigma^{2}\equiv \sigma_{\alpha\beta}
\sigma^{\alpha\beta}$ is then,
\[
\hspace{-1mm}\sigma^{2}=\frac{3}{4}\left[ \frac{2}{3}\theta-
e^{-f_{1}} \frac{u_{0}^{}}{(w_{0}^{}-w_{1}^{})}
\left( \Sigma \xx-w_{0}^{}\tt \right) \right]^{2}+
\left[\frac{e^{-f_{1}}}{u_{0}^{}(w_{0}^{}-w_{1}^{})}
\left( w_{0}^{}\ttt-\Sigma\xxx \right)\right]^{2}.
\]
The non-vanishing scalars of the Weyl tensor computed in the null tetrad
${\bf k}=2^{-1/2}(\bm{\theta}^{0}-\bm{\theta}^{1})$,
$\bm{l}=2^{-1/2}(\bm{\theta}^{0}+\bm{\theta}^{1})$,
${\bf m}=2^{-1/2}(\bm{\theta}^{2}+i\,\bm{\theta}^{3})$,
where $\bm{\theta}^{(\alpha)}$ are given in (\ref{cobasis}),
are \cite{KRAM}
\bea
&&\Psi_{0}+\Psi_{4}=2e^{-2f_{1}}\left[ \t\ttt+\x\xxx-A \right],
\label{psi0}\\
&&\Psi_{0}-\Psi_{4}=e^{-2f_{1}}\left[ 2\left( \t\xxx+\x\ttt \right)
-\ttt \xx -\xxx \tt \right],
\label{psi4}\\
&&\Psi_{2}=\frac{1}{12}e^{-2f_{1}}\left[ 4\left( \ttt^{2}-\xxx^{2}\right)+
2\left(X''_{1}-\ddot{T}_{1}\right)+\ddot{T}_{2}-X''_{2}\right],
\label{psi2}
\eea
where we have only used equations (\ref{AA}) from which we have isolated both
$\dttt$ and $\dxxx$ (which appear in $\Psi_{0}$ and $\Psi_{4}$).
From (\ref{psi0})-(\ref{psi2}) it follows that the solutions will be
in general (and at generic points) of Petrov type I.

\section{Full analysis of the case $m=1$ and explicit solutions}
\label{fullm1}
In this section we treat the case $m=1$ given in subsection \ref{m1}
so that $T_{a}(t)=c_{a} T(t)$. We give the equations
for each of the subcases
taking into account the results of subsection \ref{com} and then
some particular examples will be solved. Unless otherwise is stated,
throughout this section the generic term ``solutions'' will stand for
maximally $G_2$ not included in the previous
works on separable comoving coordinates \cite{WAGO,RUSE,AGGO},
that is, they will non-separable in comoving coordinates ``a priori''.
Furthermore, a relationship between $n$ and the type of equation of
state appears.

\subsection{$n=0$}
As was shown in subsection \ref{m2}, we have now
$\dot{T}_a(t)=c_a$, so that
from (\ref{emes}) and (\ref{M7M8}) of the Appendix \ref{ap:deffun}
and (\ref{Mi}) we get the values for $b_i=\{
c_1^2,\,c_2^2,\,c_3^2,\,c_1c_2,\,c_1c_3,\,c_2c_3,\,c_1c_2-2c_3^2,\,
c_2(c_1-1/2 c_2)\}$. Note that we have not used the fact that one can have
$b_7=b_8=0$, so we set $\alpha=\beta=0$ in (\ref{M7M8}) and (\ref{N7N8}).

Equations (\ref{AA}) become simply $\mbox{K}=c_2c_3$ and
\be
X''_3+X'_2 X'_3=c_2c_3,
\label{m1n0a}
\ee
while Eqs. (\ref{Ca}) provide no relations, (\ref{m7m8}) fixes $B$ and finally
(\ref{B}) reads
\be
\sum^8_{i=1}b_iN_i+N_7 N_8=-c_2 \left( c_1-\frac{1}{2} c_2\right)
(c_1c_2-2c_3^2),
\label{m1n0b}
\ee
where $N_i$ are defined in Appendix \ref{ap:deffun}. Equations (\ref{m1n0a})
and (\ref{m1n0b}) form a system of two first-order ordinary differential
equations for the three unknown functions $X'_a$. Thus, in general
the solutions
depend on an arbitrary function, allowing for several further
Ansatzs in order to find explicit solutions of this system of equations.
In particular, this freedom can be used in principle
to demand some extra property of the solutions, such as
particular equations of state. Indeed, it is possible
to find solutions with a $p=\gamma  \rho$ equation of state including
$\gamma=0$, that is, dust models. In fact, $X_a(x)=d_a X(x)+l_ax$
such that $X''(x)\neq 0$ and $p=0$ leads to a family of
dust solutions that can be generalized to give a bigger family
of algebraically general dust $G_2$ models \cite{SEVE}.
Nevertheless, these solutions have $q=3$ (number of
linearly independent functions among $\{N'_i\}$) so that they will
appear, within our scheme, when $n=3$ in the $m=2$-case after a
$t\leftrightarrow x$ change.
In this way, in order to avoid any superposition of solutions
between the different
subcases (using the $t\leftrightarrow x$, $n\leftrightarrow q$ ``symmetry'')
and to keep a `coherent' classification,
we should look only for solutions with the following values of
$q\in \{0,5,6,7,8\}$.
The rest of the cases ($q=1,2,3,4$) can be left to the study of
cases ($n=1,2,3,4,\; q=0$).
The cases with $q\geq 5$, following the reasoning given
in Section \ref{sepgen}, need at least two linearly independent
functions among $\{X_a(x)\}$. The solutions depend on the extra
assumption which closes the system (\ref{m1n0a})-(\ref{m1n0b}).

In the remaining possibility $q=0$ we have that $X'_a(x)=k_a$,
where $k_a$ are constants.
These spacetimes always admit a third Killing vector given by
\bean
&&\zeta=-k_1\Part+c_1\Parx+\frac{1}{2}y\left[k_1c_2-c_1k_2+2(k_1c_3-k_3c_1)
\right]\Pary \hspace{3cm}\\
&&\hspace{5cm}+\frac{1}{2}z\left[k_1c_2-c_1k_2-2(k_1c_3-k_3c_1)\right]\Parz.
\eean
Equations (\ref{m1n0a}) and (\ref{m1n0b}) give two relations on the
constants $c_a$ and $k_a$ which do not imply necessarily the appearance of
more isometries, thus the resulting solutions belong to the
tilted Bianchi perfect-fluid models. Nevertheless the study of
such solutions are
out of the aim of the present work, and will be omitted.

\subsection{$n=1$}
Following the arguments given in the subsection \ref{m2} and
looking at (\ref{Mi}), in this case we have: $m_1(t)=\dot{T}^2$,
$b_i=0$ and {$c_{i1}=\{c_1^2,\,c_2^2,\,c_3^2,\,c_1c_2,\,c_1c_3,\,c_2c_3,\,
c_{71},\,c_{81}\}$, where $c_{71}$ and $c_{81}$ are
two new arbitrary constants and $b_7$ and $b_8$ have
been absorved in $M_7$ and $M_8$. The equations containing $\ddot{T}$ are then
(\ref{Mi}) for $i=7,8$ and the $t$-equation in (\ref{AA}),
which read respectively
\bean
&&\left(c_1c_2-2c_3^2\right)\dot{T}^2-\left(\frac{1}{2}c_2+c_1\right)
\ddot{T}+\alpha=c_{71}\dot{T}^2, \\
&&\left(c_1-\frac{1}{2}c_2\right) \left(c_2\dot{T}^2+\ddot{T}\right)+\beta=
c_{81}\dot{T}^2,\\
&&c_3\left(c_2\dot{T}^2+\ddot{T}\right)=\mbox{K}.
\eean
These expressions together with the fact that the $c_a$ cannot vanish
simultaneously imply an equation for $T(t)$ of the form
\[
\ddot{T}=a\dot{T}^2+b,
\]
where $a$ and $b$ are constants.
Using this equation in the three previous expressions
and taking into account that the functions $1$ and $\dot{T}^2$ are
linearly independent, we get
the six following constraints for the constants:
\bea
c_{71}=c_1c_2-2c_3^2-\left(c_1+\frac{1}{2}c_2\right)a, \hspace{5mm}
c_{81}=\left(c_1-\frac{1}{2}c_2\right)(c_2+a), \hspace{5mm}c_3(c_2+a)=0
\label{cons1}\\
\alpha=\left(\frac{1}{2}c_2+c_1\right)b, \hspace{5mm}
\beta=\left(\frac{1}{2}c_2-c_1\right)b, \hspace{5mm}\mbox{K}=c_3b.
\hspace{15mm}
\label{cons2}
\eea
The only remaining equation involving
$t$-functions is (\ref{m7m8}), that reads now
${\cal C}_1\dot{T}^2+c_{71}c_{81}\dot{T}^4+B=0$,
implying ${\cal C}_1=B=0$ and
\be
c_{71}c_{81}=0.
\label{c71c81}
\ee
The equations (\ref{AA}), (\ref{Ca}) and (\ref{B}) for the
$x$-functions become respectively
\[
\dxxx+\xxx\xx=c_3 b,\hspace{1cm}
\sum^8_{i=1}c_{i1}N_i=0,\hspace{1cm}
N_7N_8=0.
\]
From this last relation and (\ref{c71c81}) it follows that only
two different cases may appear (if we neglect the solutions separable
in comoving coordinates):
(a) $c_{71}\neq 0$ $(\Leftrightarrow \{c_{81}=0,N_8=0,N_7\neq 0\})$
and (b) $c_{71}=0$ $(\Leftrightarrow \{c_{81}\neq 0,N_8\neq 0,N_7=0\})$.
The final system of equations for both subcases can
be written in the following compact normal form once we have defined
$\phi\equiv \xx$, $\psi\equiv \x-(1/2)\xx$ and $\varphi\equiv \xxx$
and after using (\ref{cons2})
\bea
&&\varphi '=c_3 b-\phi \varphi, \label{sisx1}\\
&&\psi '=\left(\frac{1}{2}c_2+c_1\right) b-\phi \psi+\frac{\epsilon}{{\cal K}}
\left( c_1\phi+c_2\psi \right)^2,\label{sisx2}\\
&&\phi '=\phi \left( \frac{1}{2}\phi+2 \psi\right)-2\varphi^2-c_2b-
\frac{1}{{\cal K}}\left( c_1\phi+c_2\psi-2c_3\varphi \right)^2,\label{sisx3}
\eea
where $\{\epsilon,{\cal K}(\neq 0)\}=\{0,c_{71}\},\{1,c_{81}\}$ for cases
(a) or (b) respectively.

In case (a) the solution of the constraints (\ref{cons1})
(now $c_{81}=0$) give two possibilities: (a1) $c_2+a\neq 0$,
implying then $c_2=2c_1\neq 0$, $c_3=0$, thus ${\cal K}=c_{71}=2c_1(c_1-a)
\neq 0$,
and (a2) $a=-c_2$, which gives ${\cal K}=c_{71}=2c_1c_2+c_2^2/2-2c_3^2\neq 0$.
For case (b), the constraints imply $c_3=0$, $a=2c_1c_2/(c_2+2c_1)$
with $c_2+2c_1\neq 0$ (as otherwise it would follow the vanishing
of the $c_a$) and ${\cal K}=c_{81}=c_2(c_1-c_2/2)(c_2+4c_1)/(c_2+2c_1)\neq 0$.
At this point it is interesting to give the conditions (\ref{cond})
explicitly, which read
\bean
\mbox{case (a)}&&c_{71}>0,\hspace{1cm} c_{71}^2 \dot{T}^2>
\left( c_1\phi+c_2\psi-2c_3\varphi \right)^2,\\
\mbox{case (b)}&&c_{81}>0,\hspace{1cm} c_{81}^2 \dot{T}^2<
\left( c_1\phi+c_2\psi-2c_3\varphi \right)^2,
\eean
thus, the definition of the ${\cal A}_E$-region (where $\rho + p>0$
holds) gives in general a proper determination of a domain in the
manifold and also a condition on the constants, which is
invariant under the change $t\leftrightarrow x$.
This last feature is important because these ``invariant'' conditions
allow us to prove some statements about the existence of solutions
under some extra conditions (kind of equation of state, for example)
that will hold also after changing $n\leftrightarrow q$. Using the
coordinate interchange,
the previous conditions
become $c_{71}>0$ and
$c_{71}^2 X'^2<\left( c_1\phi+c_2\psi-2c_3\varphi \right)^2$
(now the functions $\{\phi,\psi,\varphi\}$ are redefined with $T_a(t)$
replacing $X_a(x)$) in
the case (a) and $c_{81}>0$ and $c_{81}^2 X'^2>
\left( c_1\phi+c_2\psi-2c_3\varphi \right)^2$ in the case (b).

In the present case, imposing any further restriction such as
equations of state
may overdetermine the system of equations and give solutions
with more isometries. For instance, it can be shown that
an equation os state of the form $p=\gamma \rho$ implies $\gamma =1$
(stiff fluid), even for the comoving solutions,
as otherwise a third isometry appears acting
on the original Killing orbits, which became plane (i.e. a plane
$G_3$ on $S_2$).

The system of equations for the different cases has not been explicitly
solved in general, but some particular families have been found
under some extra restrictions on the
constants. In fact, when $b=0$ in case (a) the system can be completely
solved, although the solutions will have $q\leq 4$.
Analogously to what has been explained for the case $n=0$,
the values of $q$ for the representative solutions
of this case $n=1$ should be $\{0,1,5,6,7,8\}$.
We will present now some results concerning solutions with $q=0,1$.

\subsubsection{$q=0$}
The solutions with $q=0$ ($X'_a(x)=k_a$) correspond to the singular
points $(\varphi_0,\psi_0,\phi_0)$ of the differential system.
Let us begin with the case $c_3\neq 0$ ($\Rightarrow a=-c_2$) (case (a2)),
and determine then $b$ from (\ref{sisx1}).
In order to keep $w^{}_1>0$ we must avoid $\phi_0=0$.
Therefore, equations (\ref{sisx2}) and (\ref{sisx3}) give
\bea
&&\psi_0=\frac{c_2+2c_1}{2c_3}\varphi_0,\nonumber\\
&&\left(c_3\phi_0-c_2\varphi_0\right)\left[ \left(2c_1^2-c_{71}\right)c_3
\phi_0-\left(2c_1^2+c_{71}\right)c_2\varphi_0 \right]=0,
\label{varphi}
\eea
where (see above for case (a2)) $c_{71}=2c_1c_2+c_2^2/2-2c_3^2$.
These solutions do not admit a third isometry nor a
barotropic equation of state in general. In fact, it can be
shown that the only barotropic equation of state that these
maximally $G_2$ solutions can have is $\rho=p$, and those correspond to the
cases with $c_3\phi_0-c_2\varphi_0=0$. 
Let us focus our attention into the rest of the cases.
From (\ref{varphi}), and taking into account that
neither $2c_1^2+c_{71}$ nor $c_2$ can vanish in order to keep $c_{71}>0$,
we determine $\varphi_0$, so finally we have
\be
\psi_0=\left(\frac{c_2+2c_1}{2c_2}\right)Q
\phi_0,\hspace{1cm}
\varphi_0=\frac{c_3}{c_2}Q
\phi_0,
\label{psiphi}
\ee
where we have defined $Q\equiv (2c_1^2-c_{71})/(2c_1^2+c_{71})$
($Q<1$).

At this stage we can compute $p-\rho$ and realize that it is positive
everywhere, so the dominant energy condition cannot hold in the
perfect-fluid region. Nevertheless, it is important to remark here that,
within our treatment of the problem, solutions not satisfying energy (or other)
conditions may still be relevant because these conditions are {\it not} 
invariant under the interchange symmetry $t \leftrightarrow x$. Thus, we
must always check the physical properties both for the explicitly obtained
solution {\it as well as} for its partner solution with $t$ and $x$
interchanged. Otherwise, the full method will not be coherent. An illustrative
example is given, in fact, by the solutions above, because the new
solutions obtained by means of the coordinate interchange are well-behaved.
In order to see this, let us perform the change $t \leftrightarrow x$ so that
the line-element becomes
\bea
&&\wh{ds}^2=e^{\left(2\psi_0+\phi_0\right)t+2c_1X(x)}
\left(-\d t^2+\d x^2\right)+
\hspace{5cm}\nonumber\\
&&\hspace{3cm}e^{\phi_0 t+c_2 X(x)}\left(e^{2\left( \varphi_0 t +c_3 X(x)
\right)}\d y^2+e^{-2\left( \varphi_0 t +c_3 X(x)\right)}\d z^2\right),
\label{ds2n1a}
\eea
where $\psi_0$ and $\varphi_0$ are given by (\ref{psiphi}) and
the function $X(x)$ satisfies the equation
\[
X''+c_2X'^2=\frac{Q}{c_2}\,\phi_0^2,
\]
from where four cases arise:\footnote{The case $X''=0$ falls into the
previous case with $n=q=0$.}
\bean
&&\mbox{(i) }X(x)=\frac{1}{c_2}\ln \cosh \left(\phi_0\sqrt{Q}x\right),
\hspace{2cm}x\in (-\infty,\infty),\\
&&\mbox{(ii) }X(x)=\frac{1}{c_2}\ln \sinh \left(\phi_0\sqrt{Q}x\right),
\hspace{2cm}x\in (0,\infty),\\
&&\mbox{(iii) }X(x)=\frac{1}{c_2}\ln \cos \left(\phi_0\sqrt{-Q}x\right),
\hspace{1.8cm}\phi_0\sqrt{-Q}x\in (-\pi /2,\pi /2),\\
&&\mbox{(iv) }X(x)=\frac{1}{c_2}\ln (q x),
\hspace{4cm}x\in (0,\infty),\\ 
\eean
where $Q$ is positive in the first two cases, negative in the third
and $Q=0$ ($c_{71}=2c_1^2$) in the fourth.
In this last case, the solutions
admit the timelike homothetic Killing vector
$\partial / \partial t$ (not parallel to the fluid vector),
while for the corresponding `$t \leftrightarrow x$' solutions,
the homothetic Killing vector becomes $\partial / \partial x$.
The solutions
do not have a barotropic equation of state unless $2\psi_0+\phi_0=0$
(implied by $c_1=0$), which gives a further isometry, so we will
demand $c_1\neq 0$ in the following. The energy density and the
pressure read then
\[
\wh{p}+\wh{\rho}=e^{-2f_1}c_{71}\left[
\left(\psi_0+\frac{1}{2}\phi_0\right)^2\frac{1}{c_1^2}-X'^2\right],\hspace{1cm}
\wh{p}-\wh{\rho}=-e^{-2f_1}\frac{2c_{71}}
{2c_1^2+c_{71}}\phi_0^2,
\]
so in the region ${\cal A}_E$ ($\wh{p}+\wh{\rho}>0$, and remembering
$c_{71}>0$),
the dominant energy conditions {\it are fulfilled}.
Note that it can be shown that in the case (i) the condition
$\wh{p}+\wh{\rho}>0$ is automatically
satisfied, so the region ${\cal A}_E$ covers the entire spacetime.
The quantities involved in the rest of the fluid quantities
read
\[
\wh{w}^{}_0=c_{71}\left(\psi_0+\frac{1}{2}\phi_0\right)^2\frac{1}{c_1^2},
\hspace{1cm}
\wh{w}^{}_1=c_{71}X'^2,\hspace{1cm}
\wh{\Sigma}=c_{71}\left(\psi_0+\frac{1}{2}\phi_0\right)\frac{1}{c_1}X'.
\]

The case when $c_3=0$ needs also $\phi_0=0$ in order
to avoid further isometries, and $\epsilon=1$ (case (b)).
After straighforward calculations we find,
as in the previous family, that $p-\rho$
is positive everywhere thus violating the dominant energy condition.
However, we can use the coordinate interchange symmetry again and luckily
we get new solutions which do fulfil the dominant energy conditions
in the perfect-fluid region. Finally, the metric is thus given by
\be
\wh{ds}^2=e^{2\psi_0t}\cos^{-q_1^2}
(q_3 x)\left( -\d t^2 + \d x^2\right) +
\cos^{q_2^2}(q_3 x)\left( e^{2\varphi_0 t}\d y^2
+e^{-2\varphi_0 t}\d z^2\right),
\label{ds2n1b}
\ee
where $x\in(-\pi/q_3,\pi/q_3)$, we need to impose
$\varphi_0^2-\psi_0^2>0$, and we have defined
$q_1^2\equiv(\varphi_0^2-\psi_0^2)/2\varphi_0^2$,
$q_2^2\equiv(\varphi_0^2-\psi_0^2)/(\varphi_0^2+\psi_0^2)$, and 
\[
q_3\equiv 2\varphi_0^2
\sqrt{\frac{2(\varphi_0^2+\psi_0^2)}{(\varphi_0^2-\psi_0^2)
(3\varphi_0^2+\psi_0^2)}}.
\]
The energy and the pressure are given by
\bean
&&\wh{p}+\wh{\rho}=\frac{q_2^2}{q_1^2}e^{-2\psi_0t}\cos^{q_1^2}(q_3 x)\left(
2\psi_0^2\tan^2(q_3 x)-\frac{1}{2}q_1^4q_3^2\right),\\
&&\wh{p}-\wh{\rho}=-q_3^2q_2^2e^{-2\psi_0t}\cos^{q_1^2}(q_3 x)
\left(\frac{2\psi_0^2}
{\varphi_0^2+\psi_0^2}\tan^2(q_3 x)+1\right),
\eean
from where it is evident that, in the perfect-fluid region
$\wh{p}+\wh{\rho}>0$,
the dominant energy condition is always satisfied, as claimed previously.
There is no barotropic equation of state in general, though.
For the rest of the fluid quantities we have
\[
\wh{w^{}_0}=2\psi_0^2\frac{q_2^2}{q_1^2}\tan^2(q_3 x),
\hspace{5mm}
\wh{w^{}_1}=\frac{1}{2}q_1^2q_2^2q_3^2,
\hspace{5mm}
\wh{\Sigma}=q_2^2q_3\psi_0\tan(q_3 x).
\]
These two families of solutions (\ref{ds2n1a}) and (\ref{ds2n1b})
with $q=0$ are algebraically general.

\subsubsection{$q=1$}
For $q=1$ we need $X_a(x)=d_a X(x)$ such that $X''(x)\neq 0$,
which is also sufficient, because the system of equations
(\ref{sisx1})--(\ref{sisx3}) implies necessarily $N_7,N_8\propto X'^2$.
The compatibilization of this system gives some relations on the
constants: first, we have that $\epsilon=0$, so that
the solutions must belong to case (a), in order to avoid
comoving coordinates.
For $b=0$, the only maximally $G_2$ solutions without $p=\rho$ equation of
state are then given, after redefining the constants, by the
following line-element
\be
ds^2=x^\mu t^{-2 \left( 1-\mu ^2/q\right)} \left[-\d t^2+\d x^2+
x^{1-\mu}\left( x^\nu \d y^2+x^{-\nu} \d z^2 \right) \right],
\ee
where $q\equiv \mu^2+\nu^2-2\mu-1$, and the corresponding metric
after the $t\leftrightarrow x$ interchange. These solutions admit
two further conformal Killing vector fields and have no barotropic equation
of state in general. These families were found in \cite{MATHO} (pp.2320),
and we refer to this reference for further details.

For $b\neq 0$ we need $c_3\neq 0\neq d_3$ as otherwise there would
appear a third isometry, so that the solutions must belong to case (a2). Then
\[
a=-c_2, \hspace{1cm}
c_2=-\frac{d_2}{d_3}c_3, \hspace{1cm}
c_1=d_1=0,
\]
where
$d_2$ and $c_2$ cannot vanish so as to have $c_{71}>0$.
The functions $T(t)$ and $X(x)$ are given by
$\ddot{T}=d_2c_3/d_3\dot{T}^2+b$ and $X''=-d_2 X'^2+c_3b/d_3$,
and the equation of state reads $p=\rho+2d_2c_3b/d_3$.
To assure the dominant energy condition we should first
impose $b\equiv -d_3\mu^2/(d_2c_3)$,
which turns out to be also sufficient (in the ${\cal A}_E$-region).
The coordinate interchange $t\leftrightarrow x$ gives no further
solutions for the corresponding $\wh{{\cal A}}_E$-regions
(which become, in fact, identical to ${\cal A}_E$).
Two different families of solutions appear. The first is given by
\be
\d s^2=-\d t^2+\d x^2+\cos^{1+2\nu}(\mu x)\cosh^{1-2\nu}(\mu t)\d y^2+
\cos^{1-2\nu}(\mu x)\cosh^{1+2\nu}(\mu t)\d z^2, \label{nueva}
\ee
where $\mu $ and $\nu \geq 0$ are constants.
The conditions for the ${\cal A}_E$ region give $\nu \leq 1/2$ and
\[
\tanh^2(\mu t)>\tan^2(\mu x),
\]
so that we choose $t\geq 0$ and $x\in (-\pi/4\mu,\pi/4\mu)$.
The equation of state and the fluid quantities are given by
\[
\rho=\mu ^2\left[\left(\frac{1}{4}-\nu^2\right)\left(\tanh^2(\mu t)
-\tan^2(\mu x)\right)
+1\right], \hspace{1cm} p= \rho-2\mu^2,
\]
\[
u^0=\frac{\tanh(\mu t)}{\sqrt{\tanh^2(\mu t)-\tan^2(\mu x)}},\hspace{1cm}
u^1=-\frac{\tan(\mu x)}{\sqrt{\tanh^2(\mu t)-\tan^2(\mu x)}},
\]
\[
\acc_1= -\mu (u^0 u^1)^2 \tan^{-1}(\mu x) (\tanh^2(\mu t)-\tan^2(\mu x)-2),
\hspace{1cm}
\acc_0=\frac{\tan(\mu x)}{\tanh(\mu t)}\acc_1,
\]
\[
\theta=\mu\frac{(u^0)^2+(u^1)^2}{\sqrt{\tanh^2(\mu t)-\tan^2(\mu x)}}
\left(\tanh^2(\mu t)-\tan^2(\mu x)-1
\right),
\]
\bean
&&\sigma_{00}=\left(\frac{\tan(\mu x)}{\tanh(\mu t)}\right)^2 \sigma_{11},
\hspace{1cm}
\sigma_{01}=\frac{\tan(\mu x)}{\tanh(\mu t)} \sigma_{11}\\
&&\sigma_{11}=-\frac{\mu \tanh^2(\mu t)\left[(u^0)^2+(u^1)^2\right]}
{3\left(\tanh^2(\mu t)-\tan^2(\mu x)\right)^{3/2}}
\left(\tanh^2(\mu t)-\tan^2(\mu x)+2
\right)\\
&&\sigma_{22}=-\mu\nu\sqrt{\tanh^2(\mu t)-\tan^2(\mu x)}+\frac{1}
{2(u^0)^2}\sigma_{11}\\
&&\sigma_{33}=2\mu\nu\sqrt{\tanh^2(\mu t)-\tan^2(\mu x)}+\sigma_{22}.
\eean
The non-vanishing components of the Weyl tensor are
\[
\Psi_0=\Psi_4=2\nu\mu^2,\hspace{1cm}
\Psi_2=\frac{1}{2}\mu^2-\frac{1}{3}\rho ,
\]
thus the Petrov type is I for generic points.

Notice that this family of perfect-fluid solutions do not present any
curvature singularity. However, this fact may be not relevant
because, as explained above, the perfect-fluid $\cal A$-region is
extendible through the spacelike
hypersurface $\tanh^2(\mu t)-\tan^2(\mu x)=0$. A particular extension is
in fact given by the line-element (\ref{nueva}) itself when taking
every possible value of $t$ and the range $x\in (-\pi /\mu,\, \pi /\mu)$.
This particular extension is then singular, as can be immediately checked from
the above expressions.

This solution (\ref{nueva}) is in fact a good example of the power of
using non-comoving coordinates. As explained in section 2, every diagonal
$G_2$ solution can be written in comoving coordinates. Thus, we can ask
ourselves, how does solution (\ref{nueva}) look like in comoving coordinates?
In this case the change of separable coordinates to comoving coordinates
$\{t',x'\}$ can be performed explicitly and the line-element (\ref{nueva})
in the $\cal A$-region becomes
\bean
&&ds^2=\frac{1}{2}\left(t'^2+1+\sqrt{f}\right) \left\{
\frac{1}{\mu^2\sqrt{f}}\left(-\frac{1}{t'^2}\d t'^2+\frac{t'}{t'^2+1+\sqrt{f}}
\d x'^2\right)\right.\hspace{2cm}\\
&&\hspace{7cm}\left.\begin{array}{c}\mbox{ }\\\mbox{ }\end{array} 
+t'^{\,-(2\nu+1)}\d y^2+t'^{\,(2\nu-1)}\d y^2
\right\},
\eean
where $f(t',x')\equiv (t'^2-1)^2-x'^2$ and the ranges of $t'$ and 
$x'$ are restricted to $t'>0$ and $f(t',x') >0$. As we can check, it may be
difficult to find such solution if we use comoving coordinates.
Compare with its original form (\ref{nueva}), which is really {\it simple}.

The second family with $q=1$ can be obtained by simply replacing
the $\cosh (\mu t)$ functions by $\sinh(\mu t)$ everywhere (so
$\tanh (\mu t) \rightarrow \coth (\mu t)$). In this case
the range for $t$ is restricted to $t>0$ because there
is an initial spacelike singularity at $t=0$.

\subsection{$n=2$}
Despite the fact that now there are two different subcases, we begin giving
a common feature for them: we must put $c_3=0$ in order to have $\ddot{T}$
linearly independent from $1$ and $\dot{T^2}$ (see (\ref{AA})).

In the subcase (i) we take $\{1,\dot{T}^2,M_8\}$ as
linearly independent functions. To this end, due to the definition
of $M_8$,  we must also demand that $c_1-c_2/2\neq 0$.
We impose then
\[
M_7=c_{71}\dot{T}^2+c_{72}M_8,
\]
which must be an identity, i.e. multiplying this expression by $c_1-c_2/2$
we get a linear relation between the three linearly independent functions
and thus the constant coefficients must vanish, providing the following
constraints on the contants
\[
c_{71}=c_2\left(2c_1+\frac{1}{2}c_2\right), \hspace{1cm}
c_{72}=\frac{c_2+2c_1}{c_2-2c_1}, \hspace{1cm}
\alpha=c_{72}\beta.
\]
The rest of the $c_{iA}$ constants are given by
$c_{L1}=\{c_1^2,\,c_2^2,\,0,\,c_1c_2,\,0,\,0\}$,
$c_{L2}=0$, $c_{81}=0,c_{82}=1$, while $b_i=0$.
Regarding the $t$-functions, it only remains the relation (\ref{m7m8}),
which gives the equation for $T(t)$:
\bean
&&\ddot{T}^2 (2c_1+c_2)+2\ddot{T}({\cal C}_2-2\beta c_{72})
+\ddot{T}\dot{T}^2 c_2^2
+2\dot{T}^2(\frac{\beta c_2^2-2{\cal C}_1}{2c_1-c_2}-c_2{\cal C}_2)
\hspace{3cm} \\
&&\hspace{7cm}-2\dot{T}^4c_2^2c_1=4\frac{(\beta^2c_{72}
+\beta {\cal C}_2+B)}{2c_1-c_2}.
\eean
The system for the $x$-functions (\ref{AA}), (\ref{N7N8}) and
(\ref{Ca}) is given by
\bean
&&X''_3+X'_3 X'_2=0, \\
&&N_7 N_8=B, \\
&&\sum_{L=1}^6 c_{L1}N_L+c_{71}N_7={\cal C}_1,\\
&&c_{72}N_7+N_8={\cal C}_2.
\eean
This system contains four equations for three unknowns. Thus,
taking into account that we must impose $X'_3 \not\equiv 0$
to avoid a third isometry, the compatibilization of the system
gives raise to a particular family of solutions, apart from
another solutions which fall into the previous cases $n=0,1$.
After redefining some constants, the line element reads
\be
\d s^2=e^{2\mu x+2 c_1T(t)}\left( -\d t^2+\d x^2\right)+e^{c_2T(t)}\left(
e^{2\nu x}\d y^2+e^{-2\nu x}\d z^2\right),
\label{ds2m1n2}
\ee 
where $\mu$ and $\nu$ are constants, $c_2\neq 0$, $c_2+2c_1\neq 0$
and $T(t)$ satisfies
\[
\ddot{T}=\frac{1}{2(c_2+2c_1)}\left( 2 \epsilon \sqrt{\Delta}-4c_{72}\nu^2-
c_2^2\dot{T}^2\right),
\]
where $\epsilon^2=1$ and $\Delta = (c_{71}\dot{T}^2+2c_{72}\nu^2)^2+4\mu^2c_2^2
c_{72}\dot{T}^2$.
This family does not contain a barotropic equation of state except for
cases with more isometries.
The pressure, the energy density and the rest of the variables
involved in the fluid quantities are given by
\bean
&&\rho=e^{-2(\mu x+c_1T(t))}\frac{c_2}{c_2+2c_1}\left(c_{71}\dot{T}^2
- \frac{(2c_1-c_2)}{2c_2} \epsilon \sqrt{\Delta}\right),\\
&&p+\rho=e^{-2(\mu x+c_1T(t))}\frac{c_2}{c_2+2c_1}\left(c_{71}\dot{T}^2
-2c_{72}\nu^2-2\frac{c_1}{c_2}\epsilon\sqrt{\Delta}\right),\hspace{2mm}
\Sigma=c_2\mu \dot{T},\\
&&w^{}_0=\frac{1}{2}\left(c_{71}\dot{T}^2+2c_{72}\nu^2-\epsilon
\sqrt{\Delta}\right),\hspace{2mm}
w^{}_1=-\frac{1}{2c_{72}}\left( c_{71}\dot{T}^2+2c_{72}\nu^2+\epsilon
\sqrt{\Delta}  \right),
\eean
from where it is straightforward to find the pefect-fluid
quantities using the expressions given in the subsection \ref{Kine},
and realize that the region defined by $\Delta=0$ does not
represent any physical singularity but a focal zone for the
fluid congruence. The non-zero components of the Weyl tensor
read
\[
\psi_0+\psi_4=2e^{-2f_1}\mu\nu,\;
\psi_0-\psi_4=-e^{-2f_1}\nu(c_2-2c_1)\dot{T},\;
\psi_2=\frac{1}{12}e^{-2f_1}\left[(c_2-2c_1)\ddot{T}-4\nu^2\right],
\]
where $f_1=\mu x+c_1 T(t)$, so the Petrov type is I.

Although the ${\cal A}_E$ region is defined in principle by two
inequalities involving the function $T(t)$ (i.e. $\rho+p>0$, $w_0>0$),
it can be shown that after imposing the former, it suffices
the evaluation of the latter
on $\cal F$ (i.e. in $\rho+p=0$), giving then only a
condition on the constants involved.
This happens because $\Sigma$ never vanishes at $\cal F$.

For the subcase (ii) we take $\{1,\dot{T}^2,M_7\}$ to be three
linearly independent functions and impose the relation
\[
M_8=c_{81}\dot{T}^2.
\]
Following the same procedure as in the previous subcase, we must
have now $c_2+2c_1\neq 0$,
$c_2-2c_1=0$, $\beta=0$, and $c_{81}=0$, thus in fact, $M_8=0$
identically. Therefore, equation (\ref{m7m8}) has the form
\[
{\cal C}_1 \dot{T}^2+{\cal C}_2 M_7+B=0,
\]
from where it follows that ${\cal C}_1={\cal C}_2=B=0$.
The constants $c_{L1}$ and $c_{L2}$ are the same as in the subcase (i)
and also $c_{71}=0$, $c_{72}=1$.
The equation (\ref{Ca}) for $A=2$ reads then
\[
\sum_{i=1}^{8}c_{i2}N_i=N_7=0,
\]
thus $w^{}_1=0$ and therefore the possible solutions belonging
to this subcase are always separable in comoving coordinates.

To sum up this subsection, in $n=2$ there is only a family of
solutions. Its line element is given by (\ref{ds2m1n2}) and
has no barotropic equation of state.

\section{Concluding remarks}
The assumption of separability of the metric functions
in non-comoving coordinates has been shown to be a valuable tool
for the obtaining of inhomogeneous exact solutions.
By exploiting the coordinate interchange symmetry explained in section 2,
a purely mathematical classification is put forward and provides a 
systematic way of
getting new diagonal $G_2$ on $S_2$ solutions. This has been explicitly used
in the present work to construct several new solutions by means of the analysis
of the simplest case (m=1). It arises a relationship between the different
cases of the classification and some physical properties, such as
the existence of a barotropic equation of state and its explicit form.
Furthermore, the use of non-comoving coordinates shows explicitly
that the space-times obtained by the imposition of a particular Segr\'e type
of the energy-momentum tensor (for instance, perfect fluid)
may be extendible in general and with a varying algebraic type
through the extension.


\appendix
\section{The Einstein tensor and the functions $M_{i}(t)$ and $N_{i}(x)$}
\label{ap:deffun}
Here we give the explicit expressions for the Einstein tensor
for the metric given in (\ref{ds2}) computed in the natural
orthonormal co-basis (\ref{cobasis}), and the functions
$M_{i}(t)$ and $N_{i}(x)$
($i=1\ldots 8$) in terms of the metric functions defined upon equations
(\ref{expli}) and (\ref{expli2}).
The non-zero components of Einstein tensor for (\ref{ds2}) in
the frame (\ref{cobasis}) are:
\bean
&&S_{00}=e^{-2f_{1}}\left[\t \tt+\frac{1}{4}\tt^2-\ttt^2-
\dxx-\frac{3}{4}\xx^2-\xxx^2+\x \xx \right],\\
&&S_{11}=e^{-2f_{1}}\left[\x \xx+\frac{1}{4}\xx^2-\xxx^2-
\dtt-\frac{3}{4}\tt^2-\ttt^2+\t \tt \right],\\
&&S_{01}=e^{-2f_{1}}\left[\t \xx + \tt \left( \x - \frac{1}{2} \xx \right)
-2 \ttt \xxx \right],\\
&&S_{22}=e^{-2f_{1}}\left[\dx+\frac{1}{2}\dxx+\frac{1}{4}\xx^2-\xx\xxx-
\dxxx+\xxx^2 \right.\nonumber\\
&&\hspace{6cm}\left.-\dt-\frac{1}{2}\dtt-\frac{1}{4}\tt^2+\tt\ttt+\dttt-
\ttt^2\right],\\
&&S_{33}=e^{-2f_{1}}\left[\dx+\frac{1}{2}\dxx+\frac{1}{4}\xx^2+\xx\xxx+
\dxxx+\xxx^2\right.\nonumber\\
&&\hspace{6cm}\left.-\dt-\frac{1}{2}\dtt-\frac{1}{4}\tt^2-\tt\ttt-\dttt-
\ttt^2\right].
\eean
The equation (\ref{AA})
is now used to eliminate $\dttt$ and $\dxxx$ in what follows.
There is some freedom in defining the $M$ and $N$ functions, and we
could have chosen the definitions
(in particular for $M_{L}(t)$ and $N_{L}(x)$ ($L=1\ldots 6$)) keeping the $x
\leftrightarrow t$ symmetry; but, as we deal with the
$t$-functions, we define $M_{L}(t)$ as simpler as possible.
From $S_{01}^2$ we define $M_{L}(t)$ and $N_{L}(x)$ as
\be
M_{1}^{}=\t^2,\, M_{2}^{}=\tt^{2},\, M_{3}^{}=\ttt^{2},\,
M_{4}^{}=\t\tt,\, M_{5}^{}=\t\ttt,\, M_{6}^{}=\tt\ttt,
\label{emes}
\ee
\be
\begin{array}{ccc}
N_{1}^{}=-\xx^2,&N_{2}{}=-\left(\x-\frac{1}{2}\xx \right)^2,
&N_{3}^{}=-4\xxx^2,\\
N_{4}^{}=-2\xx \left(\x-\frac{1}{2}\xx \right),&N_{5}^{}=4\xx\xxx,&
N_{6}^{}=4\xxx\left(\x-\frac{1}{2}\xx \right).
\end{array}\nonumber
\ee
Similarly, from the right hand side of (\ref{ein2}), and using
the explicit expressions
for the two factors, we can define $M_{7}(t)$, $M_{8}(t)$,
$N_{7}(x)$ and $N_{8}(x)$ by means of
\bean
S_{00}+S_{22}&=&e^{-2f_{1}}\left( M_{7}(t)+N_{8}(x) \right) ,\\
S_{11}-S_{22}&=&e^{-2f_{1}}\left( M_{8}(t)+N_{7}(x) \right) ,
\eean
and such that (\ref{expli2}) holds. Thus,
\be
\begin{array}{cc}
M_{7}^{}=\t\tt-\frac{1}{2}\ddot{T}_{2}^{}-2\ttt^2-\ddot{T}_{1}^{}+\alpha,
&M_{8}^{}=\t\tt-\frac{1}{2}\ddot{T}_{2}^{}-\frac{1}{2}\tt^2+
\ddot{T}_{1}^{}+\beta,
\end{array}
\label{M7M8}
\ee
\be
\begin{array}{cc}
N_{7}^{}=\x\xx-\frac{1}{2}X_{2}''-2\xxx^2-X_{1}''-\beta,
&N_{8}^{}=\x\xx-\frac{1}{2}X_{2}''-\frac{1}{2}\xx^2+X''_{1}-\alpha,
\end{array}
\label{N7N8}
\ee
where $\alpha$ and $\beta$ are two arbitrary constants. With the help of these
constants we can always set, for instance, $M_{7}(t)+$const.$\rightarrow
M_{7}(t)$, that is, we can absorve the constants added to $M_{7},M_{8}$
into themselves. This feature is very useful to simplify the expressions
and the equations for $N_{i}(x)$.


\section{Some useful results}\label{ap:lemes}
Here we give some lemmas concerning sets of linearly independent
functions which
will be useful for relating the integers $m$ and $n$ defined
in the section \ref{sepgen}.
The first two lemmas relate a set of functions \{$f_{i}$\} ($i= 1\ldots r$),
from (an interval of) $\R$ to $\R$, with its derivatives (denoted
by a dot).
The rest give some relations of the same kind between the sets \{$f_{i}$\}
and \{$f_{i}f_{j}$\} (the combinations of products of two fuctions from
\{$f_{i}$\}). We will present an example of each type of proof
to give an outline of the procedures involved.
\begin{lema}
Let \{$f_{i}$\} be a set of $r$ linearly independent $C^{1}$
functions. Then, there are at least $r-1$ linearly independent
functions among the set \{$\dot{f}_{i}$\}.
\end{lema}
\noindent{\em Proof.\/}\hspace{3mm}Suppose that there were only $r-2$
linearly independent functions
among \{$\dot{f}_{i}$\}. This would mean that we have two independent relations
between them
\[
\sum^{r}_{i=1}a_{i}\dot{f}_{i}=0, \hspace{2cm}
\sum^{r}_{i=1}b_{i}\dot{f}_{i}=0,
\label{rel}
\]
where $a_{i}$ and $b_{i}$ are constants (or functions depending on another
independent variable). Integrating these relations we would get
\[
\sum^{r}_{i=1}a_{i}f_{i}=A\not=0, \hspace{1cm}
\sum^{r}_{i=1}b_{i}f_{i}=B\not=0
\]
where $A$ and $B$ cannot vanish. Therefore
\[
\sum^{r}_{i=1}(Ba_{i}-Ab_{i})f_{i}=0,
\]
and thus $Ba_{i}-Ab_{i}=0$ because \{$f_{i}$\} are $r$ linearly independent
functions.
But then, the two relations in (\ref{rel}) would be linearly dependent,
which is a contradiction.

\begin{lema}
\{$\dot{f}_{i}$\} is a set of $r$ linearly independent functions if
and only if \{$f_{i},1$\} is a set of $r+1$ linearly independent functions.
\end{lema}
\begin{lema}
Let $f_{1}$ and $f_{2}$ be two linearly independent functions such that
the intersection of their supports is non empty.
Then, the set \{$f_{i}f_{j}$\}=
\{$f_{1}^{2},f_{2}^{2},f_{1}^{}f_{2}^{}$\} contains three linearly independent
functions.
\end{lema}
Let us note that the condition on the supports avoids $f_{1}f_{2}\equiv 0$
(in all the interval of definition of \{$f_{i}$\}).

\noindent{\em Proof.\/}\hspace{3mm}Suppose, on the contrary, that there
existed a relation $a f_{1}^{2}+b f_{2}^{2}+c f_{1}^{} f_{2}^{}=0$. Dividing
this relation by $f_{1}^{2}$ (in its support), we would get a polynomial for
$f_{2}/f_{1}$. This would imply that $a$, $b$, and $c$ are all equal
to zero, in order to avoid the proportionality of $f^{}_1$ and $f^{}_2$.

\begin{teorem}
Let $f_{1}$, $f_{2}$, and $f_{3}$ be three linearly independent
functions such that the
intersection of any two of their supports is non empty. Then, the set
\{$f_{i}f_{j}$\}
contains at least five linearly independent functions.
\end{teorem}

\section*{Acknowledgments}
R. Vera wishes to thank the {\em Direcci\'o General de Recerca,
Generalitat de Catalunya,} for financial support.

\end{document}